\documentclass[11pt,a4paper]{article}
\usepackage{amsmath}
\usepackage{amsfonts}
\usepackage{amssymb}
\usepackage{cite}

\title{\large{Shortly on Genuine Significance of Uncertainty Relations}}
\author{\normalsize{Spiridon  Dumitru}\\ 
\normalsize{(Retired)Department of Physics, \textit{"Transilvania"} University,}\\
\normalsize{B-dul Eroilor 29, 500036 Brasov, Romania} \\ \normalsize{E-mail: s.dumitru42@yahoo.com}}
\date{\today}

\begin{document}

\maketitle
\begin{flushright}
 {\textsc{motto:}" uncertainty relations in their present form\\ 
         will not survive in the physics of future"\\
P.A.M. Dirac , 1963}
\end{flushright}
\begin{abstract}
It brings into attention briefly the genuine   significance of uncertainty  relations  and of their extrapolations for which conventional (usual) doctrine promotes unjustified ideas.  

\end{abstract}
\begin{flushleft}
Genuine Significance

\end{flushleft} 

\section{Introduction}

The Uncertainty Relations (UR) enjoy a considerable popularity, due in a large measure to the Conventional Interpretation of UR (CIUR) doctrine. The mentioned doctrine (or dogma), initiated by Copenhagen School supporters, is frequently associated with the idea (which persist so far) that UR have crucial significance in physics (for a list of relevant references see \cite{1,2,3,4,5,6}).  
          The alluded doctrine and idea were born from a core whose more important assertions are itemized in the next section.  But as we will show in Section 3 all the mentioned assertions prove themselves to be mere groundlessness statements. For atonement of the mentioned statements in Section 4 are bring into attention the genuine   significances of aboriginal formulas of CIUR formulas and a natural approach of the QMS description.   
     Many publications promote extrapolations that exceed the alluded core of CIUR. In Section 5 are presented few such extrapolations and counterarguments  concerning them.
    The Section 6 summarizes the views of the article with the conclusions that in fact the UR have not any crucial significance for physics. Such a conclusion consolidates the Dirac's intuitional guess that "\textit{uncertainty relations . . .  will not survive in the physics of future}".

\section{The main assertions of CIUR doctrine}
    A detailed examination of the mainstream publications regarding the foundations and interpretation of Quantum Mechanics (QM) show that the core of CIUR doctrine can be itemized through the following Assertions ($\textbf{\textit{A}}$): 
 \\
 
 $\bullet$ $\textbf{\textit{A}}_1$ : In an experimental reading (initiated by Heisenberg) the CIUR wants to offer an unique and generic interpretation for the \textit{thought-experimental} (\textit{te}) formula
\begin{equation}\label{eq:1}
\Delta _{te} A \cdot \Delta _{te} B \ge \hbar 
\end{equation}                                                                                                 where $ A$  and  $B$ represent   two  conjugated    observables while $\Delta_{te} A$    and $\Delta_{te} B$      denote   the            
\textit{ te}-measuring uncertainties in simultaneous measurements of $ A$  and  $B$. The mentioned formula is attributed to the "disturbing  effect" due to the (observer) measuring devices.$\blacksquare$

$\bullet$ $  \textbf{\textit{A}}_2$ : Within theoretical approach (put forward by Robertson and Schrodinger) CIUR formula \eqref{eq:1} is consolidated by QM theoretical relation
 \begin{equation}\label{eq:2}
\Delta _\psi  A \cdot \Delta _\psi  B \ge \frac{1}{2}\left| {\left\langle {\left[ {\hat A,\hat B} \right]} \right\rangle _\psi  } \right|
\end{equation}
(where the notations are the usual ones from usual QM - see also \cite{1,6}). So the uncertainties $\Delta_\psi A$  and  $\Delta_\psi B$ of two QM  observables  $ A$  and  $B$ are mutually interconnected or not as their operators $\hat A$  and $\hat B$   commute or no.$\blacksquare$

$\bullet$ $ \textbf{\textit{A}}_3$: Sustained by the mentioned meanings for the couples  $\Delta_{te} A$    and $\Delta_{te} B$     respectively   $\Delta_\psi A$  and  $\Delta_\psi B$  the relations \eqref{eq:1} and\eqref{eq:2}   are named UR and within  CIUR doctrine they  are supposed to be  connected with the description of uncertainties (errors) specific for Quantum Measurements (QMS), without any similarity in Classical Physics (CP).$\blacksquare$

$\bullet$ $ \textbf{\textit{A}}_4$: As an essential piece in aboriginal relations  (1) and (2) of CIUR, the Planck's constant $\hbar$ is regarded by CIUR as being  exclusively a QM symbol without any kind of analogue in CP.$\blacksquare$

\section{Groundlessness of the assertions $\textbf{\textit{A}}_1$ - $\textbf{\textit{A}}_4$ }
It is possible \cite{5,6}  to prove  that all basic assertions $\textbf{\textit{A}}_1$ - $\textbf{\textit{A}}_4$ of CIUR doctrine are prejudiced by insurmountable shortcomings. The alluded   shortcomings are revealed by a number of Reasons ($\textbf{\textit{R}}$) such as:
\\

$\bullet$ $\textbf{\textit{R}}_1$ : The assertion $\textbf{\textit{A}}_1$ is mere provisional fiction without any durable physical significance. This because the $\textit{te}$-formula \eqref{eq:1}  has only a transitory/temporary character, it being founded  on old resolution criterion from optics (introduced by Abe and Rayleigh). In its essence through the respective criterion  a measurement is regarded as a single sampling for the value of the measured observable. Moreover  the mentioned old  criterion was surpassed (for references see \cite{5,6} by the so called super-resolution techniques worked out in modern experimental physics. Then, instead of assertions $\textbf{\textit{A}}_1$ it is possible \cite{5,6} to appeal to some improved super-resolution ideas and formulas   able to invalidate in its very essence the respective assertion .$\blacksquare$

$\bullet$ $\textbf{\textit{R}}_2$:  Relation \eqref{eq:2} is only a restricted consequence of a Cauchy-Schwarz formula applied in QM theoretical framework. So regarded the alluded relation is nothing but a simple correlation formula   \cite{5,6}   with some ones find out in fluctuations theory  from classical statistical physics. Moreover \cite{5,6}   the respective relation is not applicable  correctly in cases of some "rebellious" situations (such are the cases with  pairs of observables :   (a) angular momentum - azimuthal angle (e.g. for examples of an atomic electron respectively of torsion pendulum), (b) number - phase (for a QM linear oscillator), (c) energy - time, (d) Cartesian momenta $p_x$ and  $p_y$ for a particle in a 2D potential well, (e) the cases with eigen states of one observable   or        (f) deficient CIUR  generalization of formulas (2) to the macroscopic quantum statistical  systems). Then it is groundlessness to regard UR as a principle (like in \cite{3}). This because a principle should be applicable without exception for all cases of a class of similar situations.$\blacksquare$

$\bullet$ $\textbf{\textit{R}}_3$:  In its essence a quantum system (e.g. an atomic electron and a linear or torsional oscillator) is endowed with random characteristics. Consequently a measurement on such a system must be conceived not as a single sampling but as a representative statistical ensemble of samplings. On the other hand the QM  theoretical framework (which incorporate relation \eqref{eq:2} via a Cauchy-Schwarz formula) deals with theoretical concepts and models about the intrinsic (inner) 
properties of the considered system but not with aspects of the measurements performed on the respective system. Moreover the measuring uncertainties regarding  an  observable A can be changed through the improving or worsening of experimental devices/procedures without any modification of  theoretical QM description of respective observable .$\blacksquare$

$\bullet$ $\textbf{\textit{R}}_4$:  Similarly with the cases studied in CP,  the description of QMS must be regarded and depicted as separate scientific branch distinct of usual QM (which deals with theoretical concepts and models of the studied systems). In the spirit of such a view we proposed \cite{6,7} to depict a QMS  as an  "\textit{informational input $\rightarrow$ output process}". Within such a process the quantum operators (associated to observables) preserve their mathematical expressions given in usual QM. But in the same process the quantum randomness of the measured system is depicted in terms of linear transformations of quantum probabilities carriers (i.e.  of the probabilistic densities  and currents associated with the corresponding  wave functions). The measuring errors are described through the differences between output and input values of probabilistic estimators ( such are averages, standard deviations, and correlations).$\blacksquare$

$\bullet$ $\textbf{\textit{R}}_5$:  The Planck's constant $\hbar$  besides its well-known quantum significance is endowed also \cite{5,6,8,9} with the quality of generic indicator for quantum randomness (stochasticity) - i.e. for the random characteristics of QM observables. Through such a quality $\hbar$ has \cite{5,6,8,9} an authentic analogue in statistical CP. The respective analogue is the Boltzmann's constant $k_B$ which is an authentic generic indicator for thermal randomness. Note that, physically, the randomness of an observable is manifested through its fluctuations \cite{5,6,8,9}.$\blacksquare$ 

\section{ Genuine   significances of formulas \\ \eqref{eq:1} and \eqref{eq:2}  and of QMS description }
The above noted reasons $\textbf{\textit{R}}_1$ - $\textbf{\textit{R}}_5$  justify to note the in the below  three decisive Observations ($\textbf{\textit{0}}$) which regard the  aboriginal  formulas \eqref{eq:1} and \eqref{eq:2} of CIUR  respectively the description of QMS. So we have:
\\

$\bullet$ $\textbf{\textit{O}}_1$: Formula \eqref{eq:1} signifies nothing but an   old and obsolete resolution criterion commonly practiced in physics of early twentieth century. It has no genuine significance for acceptable descriptions of QMS .$\blacksquare$ 

$\bullet$ $\textbf{\textit{O}}_2$ : By its essence the formula \eqref{eq:2} is only a restricted theoretical QM relation. Therefore it can't be connected in a genuine way with the description of QMS. Such description must be regarded and depicted as separate scientific study distinct of usual QM.$\blacksquare$ 

$\bullet$ $\textbf{\textit{O}}_3$: In its bare and lucrative framework, the usual QM offers solely theoretical models for own
characteristics of the investigated systems (micro particles of atomic size).  In the alluded framework QM has no connection with a genuine depiction of QMS.  The description of QMS is an autonomous subject, investigable in addition to the bare theoretical structure of usual QM. We think that, to a certain extent, our above views find some support in the Bell's remark \cite{10}: "\textit{the word (measurement) has had such a damaging effect on the discussions that . . .it should be banned altogether in quantum mechanics}".$\blacksquare$ 
\\
   
   The above observations $\textbf{\textit{O}}_1$ - $\textbf{\textit{O}}_3$ point out the genuine significance of aboriginal UR \eqref{eq:1} and \eqref{eq:2}. Then the CIUR doctrine proves oneself to be indubitably in a failure situation which deprives it of necessary qualities of a valid scientific construction. Of course that such a failure regards subsequently the description of QMS.
\newpage
\section{On extrapolations of CIUR } 
The physical core of CIUR doctrine is given by the above noted Assertions $\textbf{\textit{A}}_1$ - $\textbf{\textit{A}}_4$. But in many mainstream publications about  QM appear extrapolation ($\textbf{\textit{Ep}}$ ) that exceed the framework marked by the respective core. It is noteworthy that each of the respective  extrapolations is refuted by pertinent counterarguments ($\textbf{\textit{Ca}}$). Here below we present few couples of such $\textbf{\textit{Ep}}$  and the corresponding $\textbf{\textit{Ca}}$, as follows:
\\

$\bullet$ $\textbf{\textit{Ep}}_1$ : UR are \cite {11} the expression of "\textit{the most important principle of the twentieth century physics}".$\blacksquare$ 

$\bullet$ $\textbf{\textit{Ca}}_1$ : In its essence the above  $\textbf{\textit{Ep}}_1$ proves oneself to be nothing but an unjustifiable distortion of the truths.  Such a proof results directly from the fact that in reality the UR (1) and (2) are mere provisional $\textit{te}$-relations respectively minor (and restricted) QM theoretical formulas. So it results that, in the main, UR are insignificant things comparatively with the true important principles of the 20-th century physics (such are the ones regarding Noether's theorem, mass-energy equivalence, particle-wave duality or nuclear fission). Add here the observation that $\textbf{\textit{Ep}}_1$ must be not confused with the historically certified remark that \cite{12} : UR "\textit{are probably the most controverted formulae in the whole of the theoretical physics}". With more justice the respective remark has to be regarded as accentuating the weakness of $\textbf{\textit{Ep}}_1$.$\blacksquare$ 
\\

$\bullet$ $\textbf{\textit{Ep}}_2$ : UR entail \cite{13} the existence of some "\textit{impossibility}" (or "\textit{limitative}") principles in foundational physics.$\blacksquare$ 

$\bullet$ $\textbf{\textit{Ca}}_2$ : The extrapolation $\textbf{\textit{Ep}}_2$ was reinforced and disseminated recently \cite{13} through the topic:
"\textit{What  role  does  'impossibility'  principles  or  other limits   (e.g. . . . .  Heisenberg uncertainty, . . . ) play in foundational physics . . . ?}". Affiliated oneself with the quoted topic $\textbf{\textit{Ep}}_2$ implies two readings: (i) one which hints at \textit{Measuring Limits} (\textit{ML}), respectively (ii) another one associated with the so called "\textit{Computational Limits}" (\textit{CL}). In the \textit{ML} reading $\textbf{\textit{Ep}}_2$ presumes that the QMS accuracies cannot surpass relations \eqref{eq:1} and \eqref{eq:2}. Or as we have shown the respective presumption is completely unfounded (see also \cite{5,6,8,9}) . The \textit{CL} reading of $\textbf{\textit{Ep}}_2$ seems to be associated  mainly with the Bremermann's limit (i.e. to the maximum computational speed of a self-contained system in the universe\cite{14,15}). The alluded association is built \cite{14,15} in fact on the  application of the UR \eqref{eq:2} for the couple of  observables \textit{energy - time} ( application  which \cite{5,6,8} is wrong).   That is why the mentioned association has not any justifiable value. But the search \cite{15}  of ultimate physical limits for computations remains a subject worthy to be investigated. This because, certainly, that what is ultimately permissible in practical computational progresses depends on what are the ultimate possibilities of real physical artifacts (experiences). However, from our viewpoint, appraisals of the alluded possibilities do not require any appeal to the UR.$\blacksquare$ 
\\

$\bullet$ $\textbf{\textit{Ep}}_3$:  During a QMS the wave function corresponding to the state of measured system collapses into a particular eigenfunction associated with a unique (deterministic) eigenvalue of the implied observable.
\textbf{\textit{Addenda:}} In the spirit of $\textbf{\textit{Ep}}_3$ a leading group of scientists \cite{16} put forward the questions: "\textit{whether or not the 'collapse of the wave packet' is a physical process}" and  "\textit{How can the progressive collapse of the wave function be experimentally monitored?}".$\blacksquare$ 

$\bullet$ $\textbf{\textit{Ca}}_3$: The above extrapolation $\textbf{\textit{Ep}}_3$  is inspired from the old opinion that a QMS of a quantum  observable should be regarded as a single experimental sampling (trial) which gives a unique deterministic value. But in reality such an observable is a true random variable. Then, in a theoretical framework, such a variable must be regarded as endowed with spectrum of eigenvalues. For a given quantum state/system the mentioned eigenvalues are associated with particular probabilities incorporated within the wave function of the mentioned state/system. Consequently, from an experimental perspective, a significant measurement of a quantum observable requires \cite{17} an adequate number of samplings finished through an expressive (relevant) statistical group of data/outcomes. This because for an overall rating  of a random variable a simple sampling has not any value even if a singular sampling can be regarded as a separate physical process.  The description and experimental monitoring of such a separate process in quantum context is without scientific utility, similarly to the situation of a single sampling in a classic game with a dice. In conclusion one can say \cite{17} that extrapolation $\textbf{\textit{Ep}}_3$  is an unreasonably act, without any scientific utility (theoretical or experimental).$\blacksquare$ 
\\

$\bullet$ $\textbf{\textit{Ep}}_4$ :  "\textit{The Schrodinger's cat thought experiment remains a topical touchstone for all interpretations of quantum mechanics}".
$\textbf{\textit{Note}}$: Such or similar allegations can be found in many science popularization texts, e.g. in the ones disseminated via the Internet.$\blacksquare$

$\bullet$ $\textbf{\textit{Ca}}_4$:  The essential element in the alluded Schrodinger's experiment is represented by a single decay of a radioactive atom (which, through some macroscopic machinery, kills a cat). But the individual lifetime of a single decaying atom is a random variable. That is why the mentioned killing decay is in fact a twin analogue of the above mentioned single sampling taken into account in the above extrapolation $\textbf{\textit{Ep}}_3$   regarding  the wave functions collapse. That is why the here reported extrapolation $\textbf{\textit{Ep}}_4$ is nothing but a plain fiction.$\blacksquare$ 
\\
\newpage
     The above counterarguments $\textbf{\textit{Ca}}_1$ - $\textbf{\textit{Ca}}_4$ demolish piece by piece the extrapolations $\textbf{\textit{Ep}}_1$ - $\textbf{\textit{Ep}}_4$ of CIUR and so the respective extrapolations do not offer additional arguments for extolling UR.

\section{Conclusions}
     A survey of the previous sections discloses the fact that uncertainty relations have not any crucial significance for physics. Consequently the respective relations must be disconnected from interpretation of QM and it is senseless to speak of an uncertainty principle as a foundation/corner stone of quantum philosophy. So we give forth a class of solid arguments which come to advocate and consolidate the Dirac's intuition \cite{18}: "\textit{I think one can make a safe guess that  uncertainty relations in their present form will not survive in the physics of future}"


\begin{thebibliography}{99}
\bibitem{1} G. Auletta , Foundations and interpretation of Quantum Mechanics, (World Scientific, Singapore, 2000).
\bibitem{2} A. Cabello,  arXiv: quant-ph/0012089.
\bibitem{3} Hilgevoord J., Uffink J.  Stanford Encyclopedia of Philosophy, 2006, \verb'http://plato.stanford.edu/entries/qt-uncertainty/'
\bibitem{4} P. Busch ,T. Heinonen , P. Lahti, Physics Reports, \textbf{452}, 155-176(2007)
\bibitem{5} S. Dumitru, arXiv:quant-ph/0004013v1. 
\bibitem{6} S. Dumitru, Progress in Physics $^{*)}$ , vol.\textbf{2}, issue 2,  50-68 (2008).
\bibitem{7} S. Dumitru , A. Boer,Rom. Journ. Phys., \textbf{53}, Nos. 9–10, 1111-1116 (2008).
\verb'http://www.nipne.ro/rjp/2008_53_9-10/1111_1117.pdf '  
\bibitem{8} S.Dumitru,  Progress in Physics $^{*)}$ ,  vol. \textbf{6}, issue 4, 25-29 (2010 ).
\bibitem{9} S. Dumitru,  Physics Essays, \textbf{6}, 5-20 (1993).
\bibitem{10} J.S.  Bell,   Physics World, \textbf{3}, 33-40 (1990) . 
\bibitem{11} H. Martens,  Uncertainty Principle, Ph.D. Thesis (Technical University, Eindhoven, 1991).
\bibitem{12} M. Bunge, In:Ed. By H. Pfepfer, Denken und Umdenken (zu Werk und Werkung von Werner Heisenberg),( Piper R., Munchen, 1977).
\bibitem{13} $\textbf{\textit{Ep}}_2$ was set forth in the web page of "\textit{FQXi (Foundational Questions Institute) 2009 Essay Contest: What's Ultimately Possible in Physics? (Introduction)}":
\verb'http://fqxi.org/community/forum' and \verb'http://fqxi.org/community/forum/category/31416'. 
\bibitem{14} H.J. Bremermann,  in: Ed. M.C. Yovitts et al., Self-Organizing systems ( Spartan Books, Washington, DC, 1962) 93-106.
\bibitem{15} S.  Lloyd , Nature,  \textbf{406},1047-1054 (2000) .
\bibitem{16}  G.A.D. Briggs, J.N.Butterfield, A.  Zeilinger,  Proc. Royal Soc. A, , vol.\textbf{469}, 20130299 (2013) arXiv: 1307.1310.
\bibitem{17} S. Dumitru, Progress in Physics $^{*)}$ ,Vol. \textbf{10},Issue 2, 111 - 113 (2014). 
\bibitem{18} P.A.M. Dirac,  Scientific American, 
 \textbf{208},45-53 (May 1963).\\
 - - - - -\\
 $^{*)}$ \textit{\textbf{Progress in Physics}}  is an American scientific journal, (DC, USA): ISSN 1555-5534 (print version) and ISSN 1555-5615 (online version) and papers from it are openly accessible from the site \verb'http://ptep-online.com/index_files/issues.html'
\end{thebibliography}
\end{document}